\documentclass[pdflatex,sn-mathphys-num]{sn-jnl}

\usepackage{graphicx}%
\usepackage{multirow}%
\usepackage{amsmath,amssymb,amsfonts}%
\usepackage{amsthm}%
\usepackage{mathrsfs}%
\usepackage[title]{appendix}%
\usepackage{xcolor}%
\usepackage{textcomp}%
\usepackage{manyfoot}%
\usepackage{booktabs}%
\usepackage{algorithm}%
\usepackage{algorithmicx}%
\usepackage{algpseudocode}%
\usepackage{listings}%
\usepackage{tcolorbox}

\theoremstyle{thmstyleone}%
\theoremstyle{thmstyletwo}%

\theoremstyle{thmstylethree}%

\begin{document}

\title[Article Title]{Birds of a Feather Undermine Equity: A Strategy to Align Intent and Outcome in Team-Based Learning in Higher Education}


\author*[]{\fnm{P G} \sur{Kubendran Amos}}\email{prince@nitt.edu}

%

\affil*[]{\orgdiv{Theoretical Metallurgical Group, Department of Metallurgical and Materials Engineering}, \orgname{National Institute of Technology}, \orgaddress{ \postcode{620015}, \state{Tamil Nadu}, \country{India}}}

%


\abstract{Efforts to promote equity in higher education often rely on shared intent among instructors and students. Yet, as demonstrated in this study, when students form their own teams for Team-Based Learning (TBL) tasks, they unintentionally cluster with peers of similar socio-economic backgrounds, ultimately undermining equity. This study introduces a simple strategy to facilitate equitable team formation through a quantitative reflection of students' socio-economic backgrounds and their self-perceived preparedness. When applied, the strategy yielded balanced teams and improved performance. In its absence, team compositions became skewed and class performance declined. These findings highlight a behavioural gap between intent and outcome and underscore the need for structural supports to translate equity goals into practice.}




\maketitle

\section{Introduction}\label{sec1}

A wide range of demands are placed on higher educational institutions, particularly their educators, by various stakeholders, including employers. Of these demands, none is more contrasting than the needs of industry and the expectations of society. On one hand, potential employers seek engineers who are not only technically proficient but also creative thinkers, team players, and problem solvers~\cite{carmeli2013leadership}. Educators, on the other hand, are expected to ensure equity in their higher educational institutions~\cite{clancy2007exploring}. Indeed, one cannot be prioritized at the expense of the other. While demands of the industry drive efficient technological advancement, the significance of equity spans across multiple dimensions. As higher education serves both economic and social functions, striking a balance between professional competency and social responsibility becomes a critical challenge. Given the importance of equity, several acts, including the No Child Left Behind Act~\cite{goertz2005implementing} and the Every Student Succeeds Act (ESSA)~\cite{act2015every}, have been devised and implemented to promote equity in education. However, unlike industrial needs, the description of equity is rarely definite and straightforward. Consequently, it becomes imperative to provide a proper contextual understanding.

Equity in education ensures that all students, regardless of their socio-economic background, demographic characteristics, or personal circumstances, have access to the necessary resources, opportunities, and support systems to succeed~\cite{cook2016equity}. 
It is not merely about providing equal access to education but about addressing systemic barriers that may hinder students from fully participating in and benefiting from the learning process. 
This involves creating learning environments where students are not disadvantaged due to factors ascribed at birth or in early childhood~\cite{simonova2003pursuit}. 
Moreover, true educational equity goes beyond access - it considers how the outcomes of students are shaped by their educational experiences, recognizing that differences in the learning process stemming from their background often translate into disparities in achievement and attainment~\cite{meuret2001system,saunders2017learning}. 
After all, the ultimate purpose of equity is not just to ensure access but to create conditions where students, regardless of their starting points, have comparable opportunities to achieve similar levels of success. 
This is where equity differs from equality, a distinction widely emphasized in social justice and educational policy literature. Equality largely adopts a 'one-size-fits-all' approach, while equity requires differentiated support systems to account for structural disadvantages. 
Incorporating equity in higher education classrooms plays a crucial role in fostering a sense of belonging, psychological safety, and student empowerment.
When students from diverse socio-economic backgrounds feel that their voices, perspectives, and contributions are valued and integrated into the learning process, they develop a stronger sense of ownership over their education, leading to higher engagement and motivation~\cite{itam2019diversity, shore2018inclusive}.
At its core, equity in education is a behavioural concern - one that reflects how students perceive, internalize, and act within institutional structures meant to support fairness. Beyond the classroom, equity in educational settings changes the attitude of learners toward inherent differences, ultimately laying the foundation for a just and non-discriminatory society. 

On a global scale, failure to promote equity has led to significant disparities in access to, and outcomes in, education. A recent Education for All Global Monitoring Report points out that, despite improvements over the last 15 years, there are still 58 million children out of school globally and around 100 million children who do not complete primary education (UNESCO, 2015,~\cite{benavot2015education}). The report further highlights that inequality in education has increased, with the poorest and most disadvantaged shouldering the heaviest burden. It suggests that the world's poorest children are \lq four times more likely not to go to school than the world's richest children, and five times more likely not to complete primary school\rq \thinspace~\cite{ainscow2016collaboration}. Moreover, students from low socio-economic backgrounds are twice as likely to be low performers, indicating that personal and social circumstances remain significant obstacles to achieving their educational potential~\cite{oecd2012equity}. These disparities persist into higher education, affecting not only access and retention but also everyday group dynamics, particularly in fields like engineering where collaboration and problem-solving are essential. Students from disadvantaged backgrounds often struggle to participate on equal footing in such environments.

    While equity removes barriers to achieving similar levels of success, higher education must also respond to the evolving demands of industry. Though traditional lecture-based learning is effective in delivering theoretical knowledge, it is not sufficiently equipped to cultivate the skills demanded of engineers. The inherently complex and interdisciplinary nature of engineering problems necessitates not only technical proficiency but also the ability to work effectively in teams. A powerful pedagogical strategy that imparts technical knowledge along with teamwork, problem-solving ability, and other desired skills is Team-Based Learning (TBL), an approach that has gained increasing attention in engineering and STEM education research~\cite{michaelsen2023team}. As a structured active learning technique, TBL transforms small student groups into effective learning teams. In a TBL setting, students do not passively absorb information; instead, they construct knowledge through interactive team activities, applying course concepts in a structured, collaborative environment~\cite{freeman2012adopt}. Furthermore, students take on greater responsibility for their own learning by actively participating in discussions and contributing meaningfully to their teams. When thoughtfully developed and implemented, TBL has been shown to enhance teamwork, lateral thinking, and conflict resolution, all of which are indispensable in both engineering practice and professional settings.

Among several factors, the success of Team-Based Learning (TBL) relies significantly on the strategy adopted to form teams. Team composition, if not properly monitored, has the potential to defeat the very purpose of TBL. For instance, when students from disadvantaged socio-economic backgrounds are placed in a team without adequate balance, the effectiveness of TBL is compromised. On the other hand, forming well-balanced teams ensures equity in learning within a diverse setting, ultimately achieving the demands of both industry and society.  Therefore, developing a strategy that facilitates the formation of equitable teams is vital to the comprehensive success of TBL. In this case, structuring team formation through an equity-driven approach helps bridge socio-economic learning gaps and enhances overall student outcomes. In this work, one such straightforward strategy is proposed, and the outcome of its practice is discussed. Moreover, a notable and central observation of this study is that even in classrooms where both instructors and students value equity, behavioural tendencies still lead students to form teams with peers of similar backgrounds. While it is evident that diversity can stem from various factors, the present study specifically focuses on socio-economic background as a critical determinant of equity in higher education learning environments. 

\section{Theory: Strategy for Equitable Team}\label{sec:theory}

\subsection{Understanding the diversity}

Unlike equality, implementing equity demands a proper understanding of the diversity within a given setting. In other words, equity in Team-Based Learning (TBL) cannot be achieved without understanding the diversity in the classroom. In the present strategy, relevant information regarding socio-economic diversity in a higher education classroom is gathered with the knowledge and voluntary participation of the students. This prerequisite ensures the practice of another essential societal value - \textit{transparency} - in efforts to establish equity in TBL, or education in general. Moreover, by foregrounding voluntary disclosure and informed participation, the strategy recognises that equitable practices, in addition to addressing socio-economic disparities, must accommodate the dimension of student engagement.

Correspondingly, a structured questionnaire is framed to collect this information and ascertain the degree of diversity within the classroom. An appropriate level of anonymity is maintained by requiring roll numbers instead of student names, thereby ensuring confidentiality. The questionnaire begins with a preface that clearly communicates the purpose of gathering this information. For absolute clarity, the contents of this preface are thoroughly discussed before distributing the TBL task, allowing students to make an informed decision. Besides ethical practice, this process fosters a sense of psychological safety and trust, both of which are essential for encouraging voluntary participation in equity-driven initiatives.

In the present work, the preface includes:
\begin{tcolorbox}[colback=gray!10]
One of the fundamental goals of our education system, as stated in NEP2020, is 'full equity and inclusion as the cornerstone of all educational decisions to ensure that all students are able to thrive in the education system.' Roughly translated, offering necessary support to all interested learners considering the 'disadvantages' brought by the language barrier and by the socio-economic conditions.

This form is created as an attempt to aid in developing equitable assessments (or exams). Therefore, kindly try to answer the questions with utmost honesty and sincerity.
\end{tcolorbox}

\subsection{Introducing Equity Score}

Following the preface, which informs students about the efforts to incorporate equity, a question is posed to confirm their willing participation. 
\\
This question reads:
\begin{tcolorbox}[colback=gray!10]
 Concerning Equity suggested by our NEP2020\\
\textbf{1a.} Quite honestly, I don't care much about it. I just would like to study well, finish my course and get a job. (\textbf{Equity Score - 45)}\\
\textbf{1b.} While I am preparing for myself, I also would like to help others, and get helped by others, so that everyone has an equal chance to succeed.
\end{tcolorbox}

For students who choose 1a, no further information about their socio-economic background is requested, respecting their choice. However, an \textbf{equity score of 45} is assigned to them, which will ultimately aid in forming equitable groups for Team-Based Learning (TBL). The subsequent questions are made accessible only to students who select 1b as their answer. This approach balances individual autonomy with the need for collective data, ensuring that all students, including those opting out, are considered in the broader pursuit of classroom equity.

Although social status of an individual is governed by several factors, in the present work, it is determined exclusively by their place of birth and upbringing. Correspondingly, the question posed to understand social background of the students reads:
\begin{tcolorbox}[colback=gray!10]
 The place you are from (where you grew up) is\\
\textbf{2a.} one of the few major cities in India. (\textbf{Equity Score - 05})\\
\textbf{2b.} one of the major cities in the state. (\textbf{Equity Score - 10})\\
\textbf{2c.} a small town. (\textbf{Equity Score - 15})\\
\textbf{2d.} a rural setting. (\textbf{Equity Score - 20})\\
\textbf{2e.} not comfortable to answer. 
\end{tcolorbox}
The economic status of the students are gathered through the question: 
\begin{tcolorbox}[colback=gray!10]
 Would you say, you are from a\\
\textbf{3a.} well-off family (Economically - upper class, \textbf{Equity Score - 05}).\\
\textbf{3b.} not so well-off family, but the needs are met without any difficulty\\ (Economically - upper middle class, \textbf{Equity Score - 10}),\\
\textbf{3c.} family where the needs are met only after sufficient planning\\ (Economically - middle class, \textbf{Equity Score - 15}).\\
\textbf{3d.} family where it is bit of a struggle, sometimes, to meet the needs\\ (Economically - yet-to-be middle class, \textbf{Equity Score - 20}).\\
\textbf{3e.} not comfortable to answer. 
\end{tcolorbox}
In addition to gathering information on the socio-economic background of willing students, one more question is posed to assess their self-perceived preparedness for the upcoming TBL tasks. For it could be possible that a socio-economically disadvantaged student may, in their own way, feel well-prepared to face the TBL task compared to others. Placing such a student in a group solely based on socio-economic status might compromise equity. Therefore,  alongside socio-economic background, serious consideration is extended to self-perceived readiness of the students in forming equitable teams. 
Accordingly, a question to gauge the self-perceived readiness of student is posed, and relatively higher equity points are assigned to the responses. This question reads:
\begin{tcolorbox}[colback=gray!10]
 Concerning the TBL assessment, I think\\
4a) I will be able to manage on my own (\textbf{Equity Score - 30}).\\
4b) some support from my classmates will help me get good marks. (\textbf{Equity Score - 40)}.\\
4c) I need my classmate's support to get bit more than the pass mark. (\textbf{Equity Score - 50}).
\end{tcolorbox}

The above questions were consciously framed to ensure they are not offensive in any way. Moreover, given that English is not the first language for many students - and not even the second for some -the questions were kept simple and unambiguous for clear understanding.

\subsection{Equitable Team through Equity Score}

Each answer to the questions has a specific equity score associated with it. These scores play a pivotal role in forming equitable groups for Team-Based Learning (TBL). Based on their responses, students secure an equity score that reflects both their socio-economic status and their readiness to take on the TBL assessment. These scores range from 40 to 90. \textit{ A lower equity score generally indicates students who are capable and willing to support others, while a higher score represents those who need support to perform well in TBL activities due to socio-economic disadvantages or lower self-perceived readiness. }

An equity score of 45 is assigned to students who prefer not to be part of the efforts to form equitable teams. Reasonably assuming that these students are capable of managing the tasks on their own and can be persuaded to support others when placed in a team, a score close to the minimum is assigned.

Since the questionnaire assigns an equity score to all students, including those who prefer not to be involved in the process, these scores are subsequently used to form equitable teams. An approach reflecting the Greedy Algorithm is adopted for the \textit{balanced partitioning} of students into groups based on their equity scores.

The roll numbers of students are sorted by their equity scores in descending order. From the sorted list, the students with the highest and lowest scores are placed in a team, while those with the next highest and next lowest scores form the second group. This process is repeated until each student is assigned to a team. Since equity scores reflect the socio-economic background of students, balancing out the extremes establishes equity within a team. Stated otherwise, a student with the ability and willingness to support others is placed in a group with one who needs support and is eager to accept it. Any minor imbalances in the groups can be adjusted by considering the average equity score of the teams and minimizing variance by reallocating students based on their scores. A nearly identical average score across teams suggests equity. Ultimately, by quantifying diversity through equity scores and using balanced partitioning, equitable teams are established in a socio-economically diverse classroom.

\section{Practice: Implementing the Strategy}

The present strategy for equitable Team-Based Learning (TBL) was adopted to evaluate the understanding of sophomore students in one of the most demanding fundamental courses, Metallurgical Thermodynamics and Kinetics. These students are pursuing Metallurgical and Materials Engineering at a locally renowned higher education institution in Tamil Nadu, India - National Institute of Technology, Tiruchirappalli.

In the following semester, the same students were allowed to form their own teams for TBL in another core course, Mechanical Behaviour of Materials. Similar to the previous semester, TBL was employed to assess their understanding of critical concepts. This two-semester study not only facilitated the practical implementation of the proposed strategy but also provided a comparative framework to examine its significance - or the lack thereof - when students self-select their teams.

The TBL evaluation adopted in this study served as a continuous assessment of students' understanding, wherein three sets of technical question papers were assigned to teams at regular intervals throughout the semester. This assessment carried a maximum of 70 marks and was designed to encourage collaborative problem-solving, critical thinking, and the application of recently developed artificial intelligence tools. An overview of this assessment framework can be found in REF.

\subsection{Socio-Economic Diversity}

\begin{figure}[h]
    \centering
    \includegraphics[width=\linewidth]{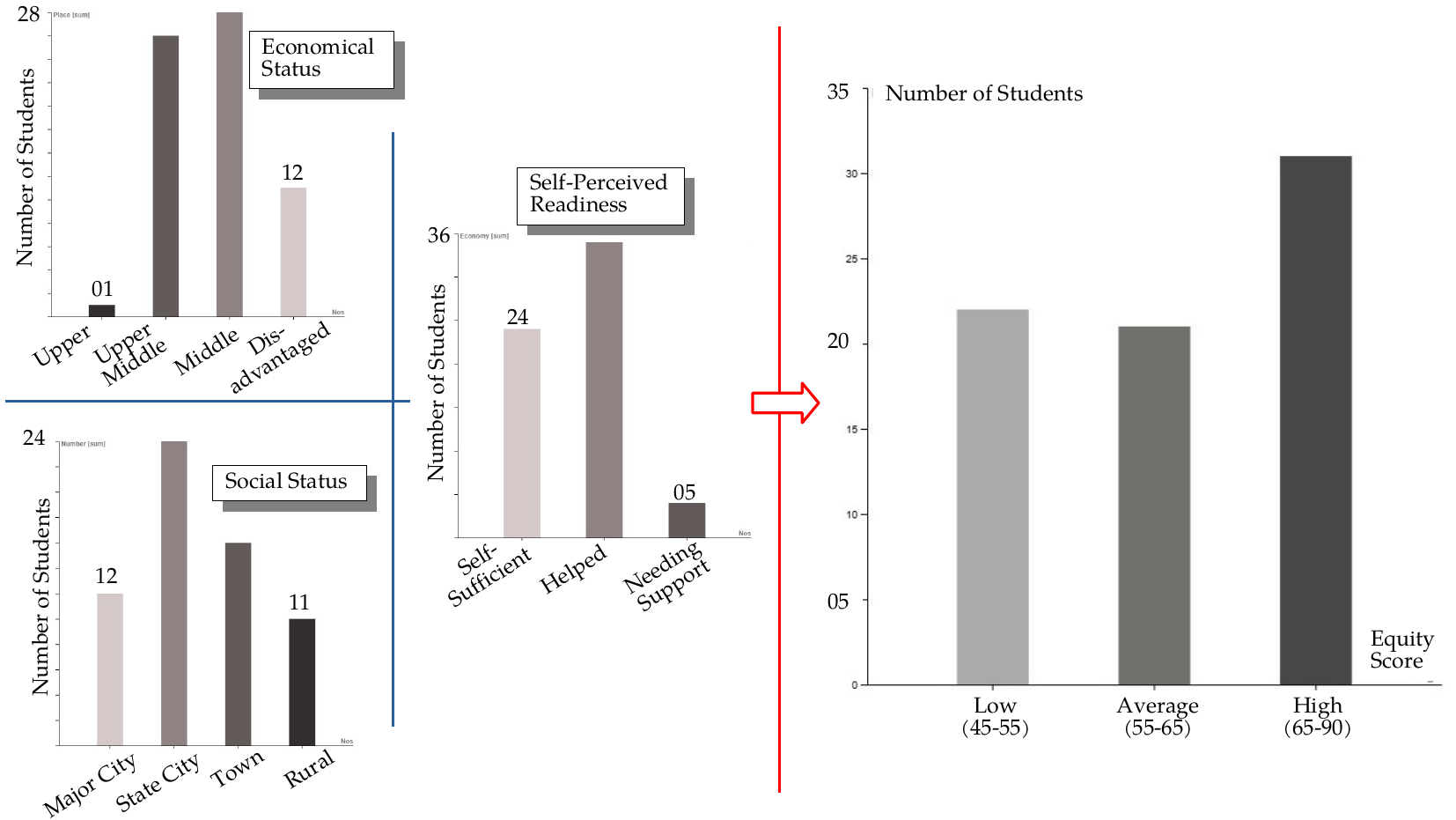} 
    \caption{\textbf{Distribution of Student Characteristics and Equity Scores}: The left panel displays bar charts representing the distribution of students across social background (major cities, state cities, towns, and rural settings), economic background (upper, upper-middle, middle, and disadvantaged), and self-perceived preparedness (self-sufficient, helped, and needing support). The right panel illustrates the distribution of students based on total equity scores, categorized as low, average, and high. High equity scores indicate students who require support from their peers to perform well in TBL assessments, while low scores represent those who perceive themselves as self-sufficient or opt out of equitable team formation.}
    \label{fig:equity_distribution}
\end{figure}

In a class of 74 students, only 9 opted out of the effort to form equitable teams for TBL. This suggests that the majority of students recognize the value of structured team formation in fostering equity within higher education learning environments. The students opting out chose the response, "Quite honestly, I don't care much about it. I just would like to study well, finish my course and get a job," in the section of the questionnaire that assessed willingness to participate in equity-driven team formation. In-keeping with the principles of participatory education (Freire, 1970), the autonomy of the students was respected, and corresponding equity scores were assigned based on their choice.

Fig.~\ref{fig:equity_distribution} illustrates the distribution of economic backgrounds among the 65 students who voluntarily disclosed their socio-economic status. A significant majority-80$\%$-belonged to the middle and upper-middle class, with 24 and 28 students, respectively. Only one student identified as economically upper-class, while 12 students (18$\%$) belonged to the lower or disadvantaged economic section.

The variation in hometowns (or places of upbringing), used in this study as an indicator of social background, is illustrated in Fig.~\ref{fig:equity_distribution}. An almost equal number of students came from major cities (12) and completely rural settings (11). Students from minor cities of the state were the most prevalent, constituting nearly 37$\%$, while 18 students identified as belonging to a small town.

Throughout the entire process of gathering relevant information, only three students (out of 130 possible responses) selected the option 'Not comfortable to answer'. In such cases, students were assigned the lowest equity score within that category, under the assumption that this particular social or economic factor did not significantly contribute to their perceived need for equitable support.

The self-perceived preparedness of students in facing the task, which is considered one of the critical factors in forming an equitable team in the present study, is shown in Fig.~\ref{fig:equity_distribution}. This illustration indicates that a majority of students - 36 of them - believe that their understanding of the subject, and ultimately the marks they secure, will be enhanced with team support. A significant portion, about 37$\%$, are confident they can complete the assessment alone. Correspondingly, they are assigned the lowest equity score in this category to prioritize those who need team support. Fig.~\ref{fig:equity_distribution} further shows that five students explicitly claim they cannot perform well on their own and require the support of their team to gain a convincing understanding of the course and achieve a decent mark in the assessment.

The overall diversity of students based on the equity score, considering social, economic, and readiness factors, is illustrated in Fig.~\ref{fig:equity_distribution}. Students with a high equity score are those from socio-economically disadvantaged backgrounds who need the support of team members to perform well or convincingly in the assessment. The low-equity score bin in Fig.~\ref{fig:equity_distribution} comprises students who perceive themselves as ready to take on the assessment independently or those who do not prefer to be part of the equitable efforts. Fig.~\ref{fig:equity_distribution} further indicates that the high equity-score group outnumbers the low and mean scores, suggesting that the majority of students could benefit from equitable TBL activities in this diverse classroom.

\subsection{Self-Perceived Readiness and Socio-Economic Background}

\begin{figure}[h]
    \centering
    \includegraphics[width=\linewidth]{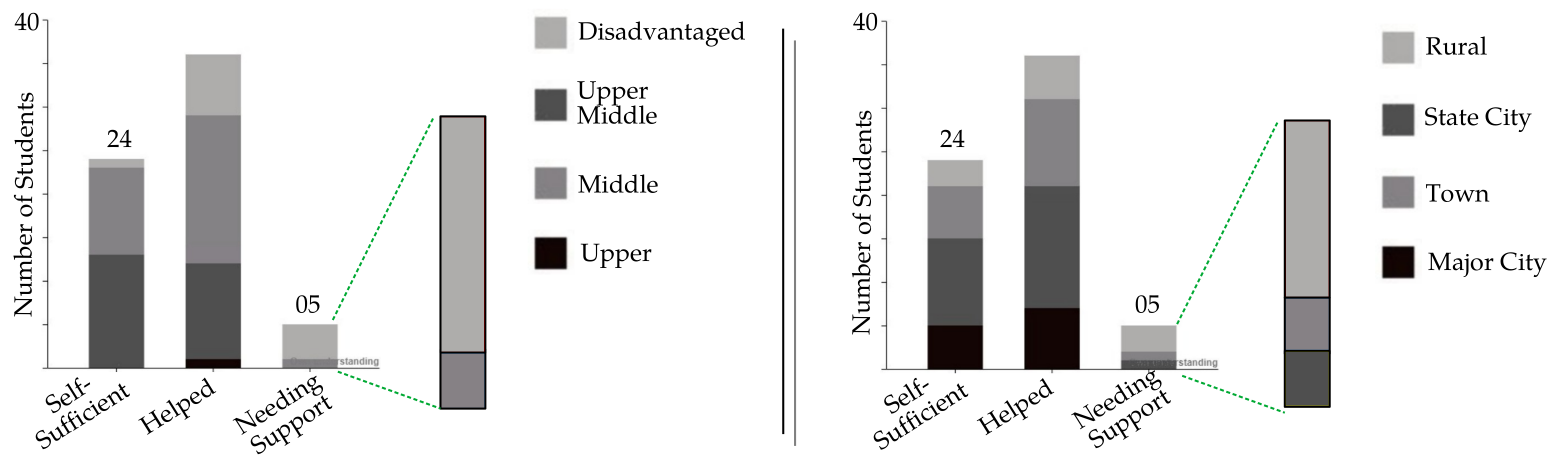} 
    \caption{The distribution of students' self-perceived preparedness - self-sufficient, helped, and needing support - varies across social (major cities, state cities, towns, rural) and economic (upper, upper-middle, middle, disadvantaged) backgrounds. Students from disadvantaged backgrounds are more likely to seek peer support, while those from privileged backgrounds tend to feel self-sufficient, highlighting the impact of socio-economic factors on learning confidence and the need for equity-driven team strategies.}
    \label{fig:SelfPerceived}
\end{figure}

The readiness of students in taking up the assessment, as illustrated in Fig.~\ref{fig:equity_distribution}, is indeed not an inherently inherited factor. However, it is interesting to examine the influence of socio-economic background on this self-perceived preparedness. Correspondingly, the socio-economic composition of students who need team assistance and those prepared to face the assessment independently is analyzed and presented in Fig.~\ref{fig:SelfPerceived}. Of the five students who claim to need team support, four come from economically disadvantaged backgrounds, and three belong to a rural area. In other words, 80$\%$ and 60$\%$ of the students who critically need team support to perform convincingly in the TBL assessment are economically and socially disadvantaged, respectively. This composition is notably reversed when considering students who are prepared to take the assessment on their own. Only 4$\%$ and 12$\%$ of those who perceive themselves as capable of managing the assessment independently are economically and socially disadvantaged. Stated otherwise, Fig.~\ref{fig:SelfPerceived} suggests that socio-economic factors influence students' readiness to face a rigorous technical task. Moreover, given the option, students from disadvantaged socio-economic backgrounds are largely eager to learn with the team. This is exactly what the present work aims to address by strategically forming equitable teams.

\subsection{Equitable vs Own Teams}

\begin{figure}[h]
    \centering
    \includegraphics[width=\linewidth]{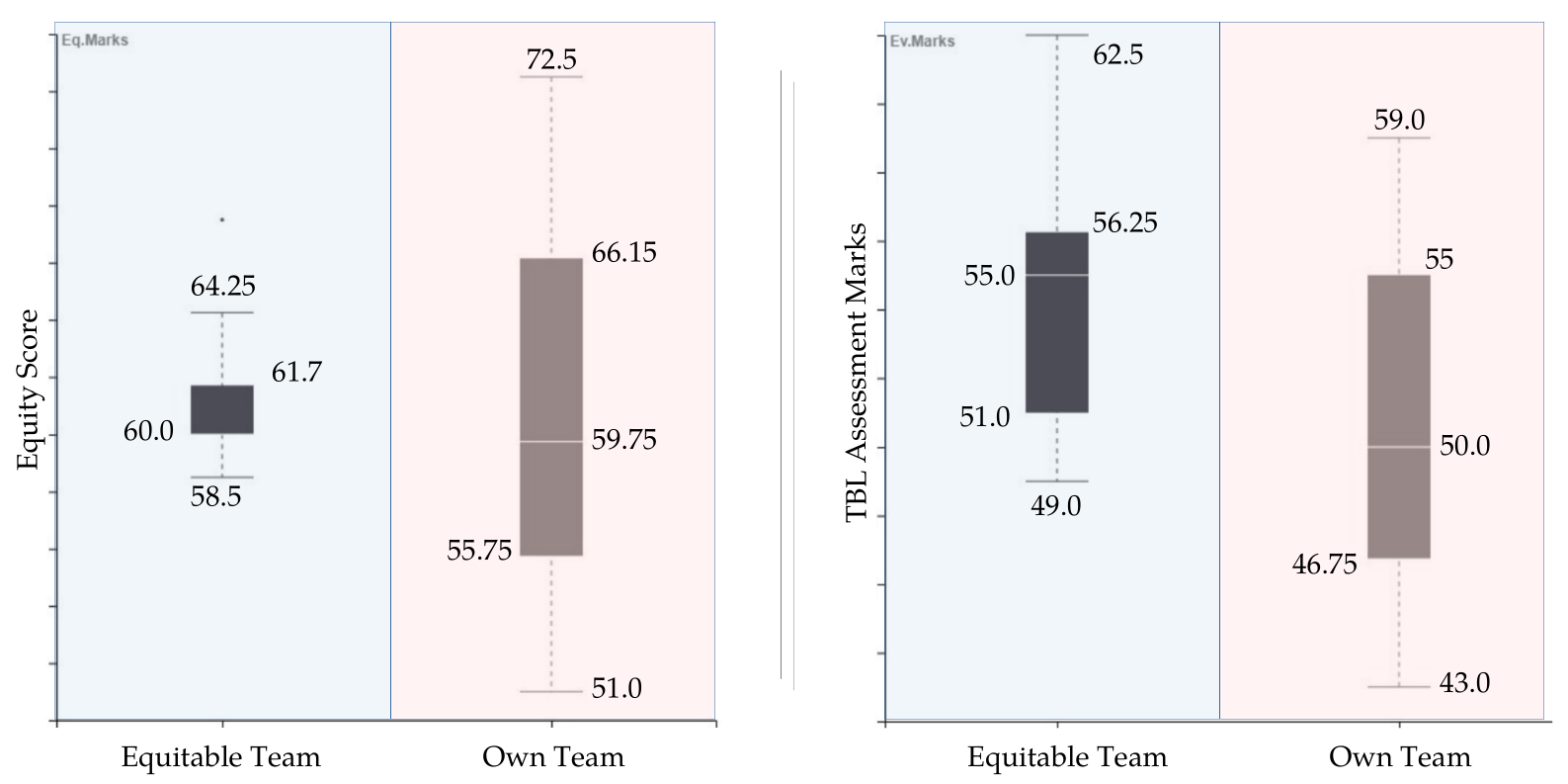} 
    \caption{Comparison of Equity Scores and TBL Assessment Marks Across Semesters. The left panel compares the equity scores of teams from the first semester (equity-driven team formation) and the second semester (self-formed teams), showing greater balance in the former. The right panel presents the TBL assessment marks for both semesters, highlighting the decline in performance when teams were self-formed.}
    \label{fig:eqMarks}
\end{figure}

In the first semester of a two-semester sophomore course, the present strategy was adopted to form equitable teams, whereas in the second semester, students were encouraged to form their own teams. To analyze the impact of this shift in team formation, the equity scores of self-formed groups were calculated based on the individual scores assigned in the previous semester. 

The equity scores of the teams from both semesters are presented as a box plot in Fig.~\ref{fig:eqMarks}. In the first semester, when the present strategy was adopted, the Greedy Algorithm-based balanced distribution of students resulted in team equity scores ranging from 58.5 to 62.25, with a median of 60. This narrow range demonstrates the effectiveness of structured team formation in maintaining equity. However, in the second semester, when students formed their own teams, the equity scores varied significantly across teams. The median equity score slightly dropped to 59.75, but more notably, the range widened from 51 to 72.5, indicating an imbalance in team composition. This disparity suggests that students with similar socio-economic backgrounds and preparedness levels clustered together, reinforcing pre-existing differences. Moreover, the imbalance in the teams reflects the well-documented behavioural tendency in which individuals, often without conscious intent, associate with others who share similar socio-economic background, ultimately perpetuating social hierarchies even within an environment previously exposed to equity.

A box-plot representation of the marks secured by the teams across the two semesters is presented in Fig.~\ref{fig:eqMarks}. In the first semester of the sophomore year, when the present strategy was adopted to form equitable teams, the maximum mark was 62 in a continued TBL assessment out of 70 marks. The minimum mark was 49, with 55 marks being the median. In the next semester, when students formed their own teams, both the maximum and minimum marks dropped to 59 and 43, respectively. Moreover, the median lowered from 55 to 50 when the assessment was undertaken by self-formed teams. The box plots in Fig.~\ref{fig:eqMarks} indicate that the implementation of the strategy in the present study, in addition to forming equitable teams, visibly enhances the overall performance of the class.

\subsection{Equity vs Assessment Scores}

\begin{figure}[h]
    \centering
    \includegraphics[width=\linewidth]{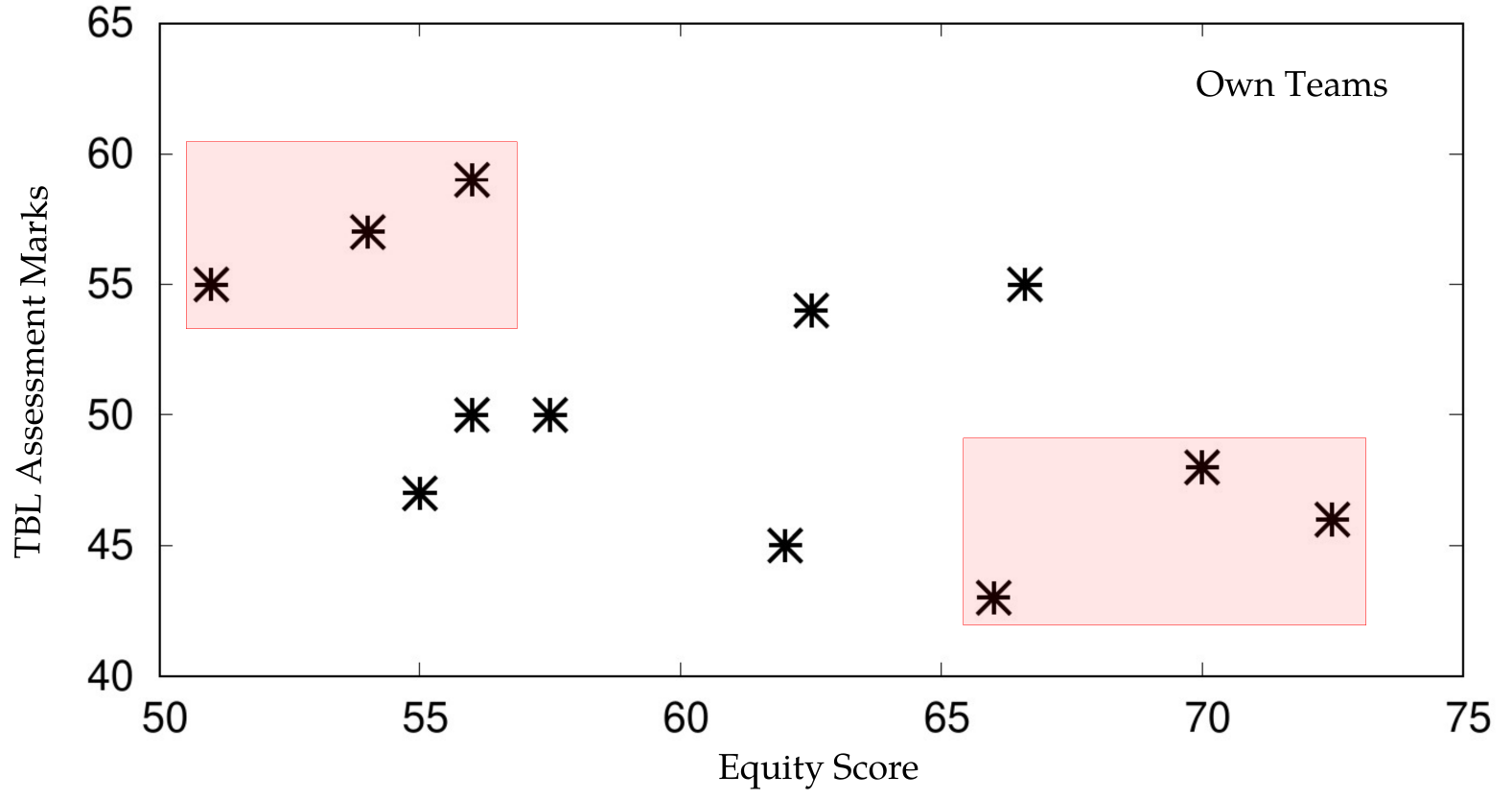} 
    \caption{Relationship Between Equity Scores and TBL Assessment Marks in Self-Formed Teams. Equity scores compared with  TBL assessment marks in the self-formed teams of the second semester. Two contrasting regions are highlighted: low equity score teams securing high marks and high equity score teams securing low marks. The pattern suggests that students who required support underperformed due to the lack of structured team formation, while self-sufficient students excelled, reinforcing the need for equity-driven team strategies in collaborative learning.}
    \label{fig:eqVSmarks}
\end{figure}

In Fig.~\ref{fig:eqVSmarks}, the equity scores of self-formed groups are plotted against the marks secured by the teams. While no distinct trend emerges, the highlighted sections indicate that teams with higher equity scores tend to secure lower marks, whereas those with lower equity scores perform better. This suggests that when students were allowed to form their own teams, those who required support struggled due to its absence, while students who were more self-sufficient excelled without additional assistance. The decline in overall class performance can thus be linked to the lack of structured support for students with high equity scores and the absence of motivation for those with lower scores. This imbalance in team composition ultimately undermines the effectiveness of an otherwise efficient Team-Based Learning (TBL) approach, reinforcing the necessity of equity-driven strategies in active learning environments.

\section{Conclusion}

The findings of this study establish that an equity-driven strategy for team formation enhances learning outcomes and mitigates socio-economic disparities in Team-Based Learning (TBL). The proposed approach, grounded in quantitative equity scoring and the Greedy Algorithm-based balancing method, ensures that students with varying socio-economic backgrounds and self-perceived preparedness collaborate effectively. When implemented, this strategy results in balanced team compositions and improved academic performance, confirming that equity-focused interventions are critical in active learning environments.

The comparative analysis between equity-driven and self-formed teams reveals that in the absence of intervention, students naturally cluster with peers of  socio-economic status, often unconsciously,, reinforcing disparities rather than fostering inclusivity. Performance variability increased, overall class performance declined, and students from disadvantaged backgrounds, particularly those with lower self-perceived preparedness, struggled to access peer support, affecting their engagement and outcomes. A notable observation is that most students demonstrated willingness and interest in forming equitable groups, provided that the mechanisms to facilitate such formations were in place. This highlights that while students recognize the value of equitable collaboration, they require structured provisions to translate their intent into effective team-based learning experiences. Furthermore, behavioural patterns like homophily in team formation or excessive perception of readiness can unintentionally subvert equity goals, even in well-intentioned environments. By embedding equity into the design of team formation, the proposed strategy acts as a behavioural scaffold that aligns group dynamics with the broader objectives of inclusion and fairness.

While the study is limited in scale, being conducted within a single institution and academic discipline, it provides empirical evidence supporting equity-focused team formation, it is limited in scale, being conducted within a single institution and discipline. A broader study encompassing larger student cohorts, multiple institutions, and different academic fields would enhance the generalizability of the findings. Additionally, the strategy focuses exclusively on socio-economic background and self-perceived preparedness, without considering other critical dimensions of equity such as gender, race, and neurodiversity, which play a crucial role in educational inclusivity. In future work, we intend to expand the scope of equity factors and examine the proposed strategy across broader educational contexts, with the goal of developing a more comprehensive behavioural framework for ensuring fair and effective team-based learning in diverse classroom settings.


\end{document}